\documentclass[aps,pra,twocolumn,10pt,notitlepage,showpacs]{revtex4-1}

\usepackage{braket}
\usepackage{amsmath}
\usepackage{color}
\usepackage{graphicx}
\usepackage{subcaption}
\usepackage{caption}
\usepackage{amssymb}
\usepackage{amsfonts}
\usepackage[parfill]{parskip}
\usepackage{hyperref}

\def \lbra {\langle}

\newtheorem{Lemma}{Lemma}

\newcommand{\proof}{\noindent {\bf Proof: }}
\newcommand{\qed}{$\Box$}
\usepackage[format=plain,justification=centerlast,singlelinecheck=true]{caption}

\begin{document}


\title{Spatial search by quantum walk is optimal for almost all graphs}


\author{Shantanav Chakraborty$^{1,2,*}$}  
\author{Leonardo Novo$^{1,2,*}$}
\author{Andris Ambainis$^3$}
\author{Yasser Omar$^{1,2}$}
\affiliation{$^1$Physics of Information and Quantum Technologies Group, Instituto de Telecomunica\c{c}\~oes, Portugal}
\affiliation{$^2$Instituto Superior T\'{e}cnico, Universidade de Lisboa, Portugal}
\affiliation{$^3$Faculty of Computing, University of Latvia}
\collaboration{$^*$both authors have equal contribution}
\begin{abstract}
The problem of finding a marked node in a graph can be solved by the spatial search algorithm based on continuous-time quantum walks (CTQW). However, this algorithm is known to run in optimal time only for a handful of graphs. In this work, we prove that for Erd\"os-Renyi random graphs, i.e.\ graphs of $n$ vertices where each edge exists with probability $p$, search by CTQW is \textit{almost surely} optimal as long as $p\geq \log^{3/2}(n)/n$. Consequently, we show that quantum spatial search is in fact optimal for \emph{almost all} graphs, meaning that the fraction of graphs of $n$ vertices for which this optimality holds tends to one in the asymptotic limit. We obtain this result by  proving that search is optimal on graphs where the ratio between the second largest and the largest eigenvalue is bounded by a constant smaller than 1. Finally, we show that we can extend our results on search to establish high fidelity quantum communication between two arbitrary nodes of a random network of interacting qubits, namely to perform quantum state transfer, as well as entanglement generation. Our work shows that quantum information tasks typically designed for structured systems retain performance in very disordered structures.
\end{abstract}
\date{11th March, 2016}
\maketitle

Quantum walks provide a natural framework for tackling the spatial search problem of finding a marked node in a graph of $n$ vertices. In the original work of Childs and Goldstone \cite{Childs_spatial_search}, it was shown that continuous-time quantum walks can search on complete graphs, hypercubes and lattices of dimension larger than four in $\mathcal{O}{(\sqrt{n})}$ time, which is optimal. More recently, new instances of graphs have been found where spatial search works optimally. These examples show that global symmetry, regularity and high connectivity are not necessary for the optimality of the algorithm  \cite{Meyer_symmetry,dimred,meyer2}. However, it is not known how general is the class of graphs for which spatial search by quantum walk is optimal.\\
Here we address the following question: If one picks at random a graph from the set of all graphs of $n$ nodes, can one find a marked node in optimal time using quantum walks? We show that the answer is \emph{almost surely} yes. Moreover, we adapt the spatial search algorithm to protocols, for state transfer and entanglement generation between arbitrary nodes of a network of interacting qubits, that work with high fidelity for \emph{almost all} graphs, for large $n$ (\textit{nodes} and \textit{vertices} are used interchangeably throughout the paper). Thus, besides showing that spatial search by quantum walk is optimal in a very general scenario, we also show that other important quantum information tasks, typically designed for ordered systems, can be accomplished efficiently in very disordered structures.\\
We obtain our results by studying the spatial search problem in Erd\"{o}s-Renyi random graphs, i.e.\ graphs of $n$ vertices where an edge between any two vertices exists with probability $p$ independently of all other edges, typically denoted as $G(n,p)$ \citep{ER59, ER60}. Note that our approach is different from the quantum random networks of non-interacting qubits defined in  \citep{perseguers2010quantum}, where two nodes are connected if they share a maximally entangled state, having in view long-distance quantum communication. Also, in Refs.~\citep{QCN2,QCN1}, the authors compare the dynamics of classical and quantum walks on  Erd\"{o}s-Renyi graphs and other complex networks, although with a different perspective from our work.\\
In our work, we show that search is optimal on $G(n,p)$ with probability that tends to one as $n$ tends to infinity, as long as $p\geq\log^{3/2}(n)/n$. It can be demonstrated that when $p=1/2$, $G(n,1/2)$ is a graph picked at random from the set of all graphs of $n$ nodes in an unbiased way, i.e. each graph is picked with equal probability. This allows us to conclude that spatial search by quantum walk is optimal for almost all graphs from this set. To obtain this result, we prove a sufficient condition regarding the adjacency matrix of graphs where search is optimal: the eigenstate corresponding to its largest eigenvalue must be  sufficiently delocalized and the ratio between the second largest and the largest eigenvalues must be bounded by a constant smaller than 1.\\
This general result also allows us to prove that search is optimal for graphs sampled uniformly from the set of all regular graphs, also known as random regular graphs. Thus, this leads us to conclude that spatial search by quantum walk is optimal for almost all regular graphs.\\
\textbf{\emph{A sufficient condition for optimal quantum search --}}
Let $G$ be a graph with a set of vertices $V=\{1,\dots,n\}$. We consider the Hilbert space spanned by the localized quantum states at the vertices of the graph $\mathcal{H}=\text{span}\{\ket{1},\dots, \ket{n}\}$, and the following search Hamiltonian
\begin{equation}\label{search_ham}
H_{G}=-\ket{w}\bra{w}-\gamma A_{G},
\end{equation}
where $\ket{w}$ corresponds to the solution of the search problem, $\gamma$ is a real number and $A_{G}$ is the adjacency matrix of a graph $G$ \cite{Childs_spatial_search}. We say that quantum search by continuous time quantum walk is optimal on a graph $G$ if there is an initial state $\ket{\psi_0}$, irrespective of $w$, and a value of $\gamma$ such that after a time $T=\mathcal{O}(\sqrt{n})$  \cite{Farhi_analog_grover}, the probability of finding the solution upon a measurement in the vertex basis is $|\braket{w|e^{-iH_{G}t}|\psi_0}|^2=\mathcal{O}(1)$. The initial state $\ket{\psi_0}$ is usually chosen to be the equal superposition of all vertices, i.e.\ the state $\ket{s}=\sum_{i=1}^n \ket{i}/\sqrt{n}$, since it is not biased towards any vertex of the graph. We start by proving the following general lemma regarding the spectral properties of $A_{G}$ and the optimality of search: 
\begin{Lemma}\label{lemma_search}
Let $H_1$ be a Hamiltonian with eigenvalues $\lambda_1 \geq \lambda_2 \geq 
\ldots \geq \lambda_k$ (satisfying $\lambda_1=1$ and $|\lambda_i|\leq c < 1$ for all 
$i>1$) and eigenvectors $\ket{v_1}=\ket{s}$, $\ket{v_2}, \ldots, \ket{v_k}$ and 
let $H_2=\ket{w}\bra{w}$ with $|\lbra w \ket{s}|=\epsilon$. 
For an appropriate choice of $r=O(1)$, applying the Hamiltonian $(1+r) H_1+H_2$ to 
the starting state $\ket{v_1}=\ket{s}$ for time $\Theta(1/\epsilon)$ results in a state $\ket{f}$ with $|\lbra w \ket{f}|^2\geq \frac{1-c}{1+c}-o(1)$. 

\proof See Section I in Supplemental Material. 
\end{Lemma}
Thus, if $\lambda^A_1 \geq \lambda^A_2 \geq \ldots \geq \lambda^A_n$ are the eigenvalues of the adjacency matrix $A_G$, we choose $\gamma=1/\lambda^A_1$ and consequently, $H_1= \gamma A_G$. If $\ket{s}$ is an eigenstate of $A_G$ corresponding to its largest eigenvalue $\lambda^A_1$, and since  $|\lbra w \ket{s}|=1/\sqrt{n}$, we have that search is optimal as long as $\lambda^A_2/\lambda^A_1\leq c<1$, following Lemma \ref{lemma_search}. We will see that Erd\"{o}s-Renyi graphs and random regular graphs fulfil this property,  leading to the conclusion that search is optimal for \emph{almost all} graphs and also for \emph{almost all} regular graphs (the latter is discussed in Section II of Supplemental Material). 

In fact, Lemma \ref{lemma_search} implies that for any regular graph having a constant normalized algebraic connectivity, quantum search is optimal \footnote{ Normalized algebraic connectivity is defined as the second largest eigenvalue of the symmetric normalized Laplacian defined as $L'=D^{-1/2} LD^{-1/2}$, where $L$ is the Laplacian of the graph, with $D$ being a diagonal matrix where the $i^{th}$ diagonal entry is the degree of vertex $i$. In such a case we can define $H_1=I-L'$}. This is in contrast to Ref.~\cite{meyer2} where two examples of regular graphs \footnote{One of them is almost regular with $n-2$ vertices having degree $n/2$, while the other two having degree $n/2+1$.} with low normalized algebraic connectivity are given, such that quantum search is optimal on one and non-optimal on the other. This result showed that normalized algebraic connectivity is not a necessary condition for fast quantum search: when connectivity is low, search can be fast or slow depending on the graph. On the other hand, Lemma \ref{lemma_search} proves that high connectivity is indeed a sufficient condition.
%
\\~\textbf{\emph{Quantum search on Erd\"{o}s-Renyi random graphs -- }}
Let us consider a graph $G(n)$ with a set of vertices $V=\{1,\dots,n\}$. We restrict ourselves to simple graphs, i.e.\ graphs which do not contain self-loops or multiple edges connecting the same pair of vertices. The maximum number of edges that a simple graph $G(n)$ can have is $N={n \choose 2}$. Thus, there are ${N \choose M}$ graphs of $M$ edges and the total number of (labelled) graphs is $\sum_{M=0}^N {N \choose M}=2^N$ \cite{graph_enumeration}. Now let us consider the random graph model $G(n,p)$, a graph with $n$ vertices where we have an edge between any two vertices with probability $p$, independently of all the other edges \cite{ER59,ER60,bollobas_book}. In this model, a graph $G_0$ with $M$ edges appears with probability $P\{G(n,p)=G_0\}=p^M (1-p)^{N-M}$. In particular, if we consider the case $p=1/2$, each of the $2^N$ graphs appears with equal probability $P=2^{-N}$.
In their seminal papers, Erd\"{o}s and Renyi introduced this model of random graphs and studied the probability of a random graph to possess a certain property $Q$ \cite{ER59,ER60}. They studied properties like connectedness of the graph, the probability that a certain subgraph is present, etc. They introduced the terminology stating that \emph{almost all graphs} have a property $Q$ if the probability that a random graph $G(n,p)$ has $Q$ goes to $1$ as $n\rightarrow \infty$. Equivalently, it can be stated that $G(n,p)$ \emph{almost surely} has property $Q$. Interestingly, certain properties of random graphs arise suddenly for a certain critical probability $p=p_c$, where this probability depends typically on $n$. More precisely, if $p(n)$ grows faster than $p_c(n)$, the probability that the random graph has property $Q$
goes to $1$ in the asymptotic limit, whereas if it grows slower than $p_c(n)$ it goes to $0$. For example above the percolation threshold, i.e. when $p>\log(n)/n$ the graph is almost surely connected, whereas if $p<\log(n)/n$ the graph has almost surely isolated nodes. 

\begin{figure}[h!]
        \centering
        \begin{subfigure}[b]{0.5\textwidth}
                \includegraphics[width=0.45\textwidth]{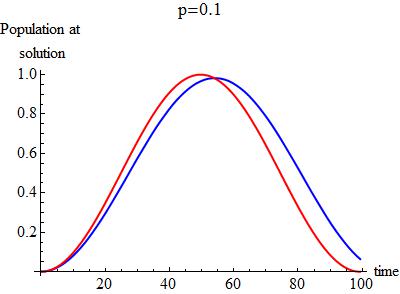}
\includegraphics[width=0.45\textwidth]{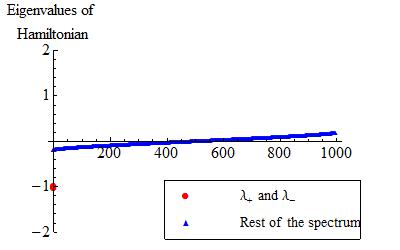}                
                \caption{~}
                \end{subfigure}%
        ~ 
        \\
        ~ 
        \begin{subfigure}[b]{0.5\textwidth}
                \includegraphics[width=0.45\textwidth]{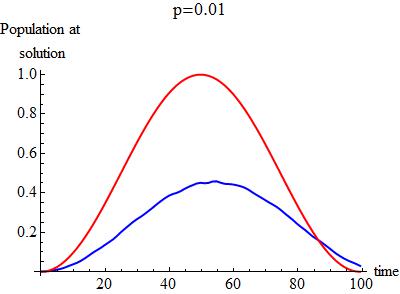}
\includegraphics[width=0.45\textwidth]{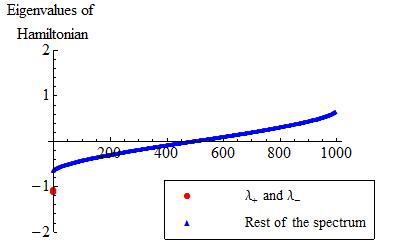}                
                \caption{~}
                \end{subfigure}\\
                \begin{subfigure}[b]{0.5\textwidth}
                \includegraphics[width=0.45\textwidth]{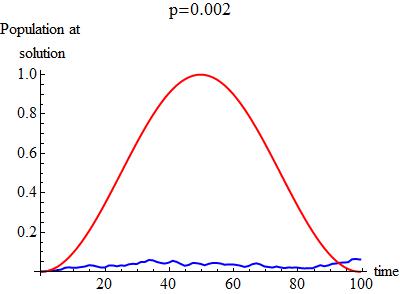}
\includegraphics[width=0.45\textwidth]{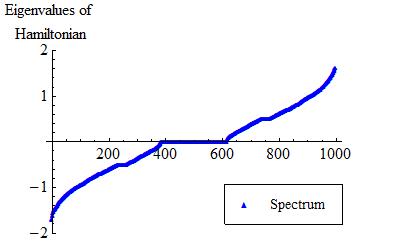}                
                \caption{~}
                \end{subfigure}
\caption{\small{Left side: probability of observing the solution calculated numerically (blue curve) compared to the prediction from Eq.~(\ref{probsucc}) (red curve), obtained in the limit $n\rightarrow \infty$, using degenerate perturbation theory. We fix the number of vertices $n=1000$ and $p=0.1, 0.01$ and $0.002$ in a), b) and c), respectively. Right side: Spectrum of the search Hamiltonian for instances of random graphs that provide the dynamics represented on the left side. In red, the two lowest eigenvalues are shown in a) and b), which are clearly isolated from the rest of the spectrum shown in blue. In c) this does not happen since $p$ is close to $1/n$, which is the percolation threshold and thus the semicircle law is not valid. We see that, the larger the gap between the two lowest eigenvalues $\lambda_{\pm}$ and the rest of the spectrum, the better is the prediction from Eq.~(\ref{probsucc}) for the probability of success. When the two lowest eigenvalues are not isolated, the probability of observing the solution is low and the algorithm does not provide speed-up with respect to classical search.}}
\label{fig:probs_spectra}
\end{figure}

In this work, we are interested in the threshold value of $p$ for which quantum search becomes optimal, i.e.\ a marked vertex from the graph can be found in $\mathcal{O}(\sqrt{n})$ time. We consider the search Hamiltonian in Eq.~(\ref{search_ham}) for Erd\"{o}s-Renyi random graphs $H_{G(n,p)}=-\ket{w}\bra{w}-\gamma A_{G(n,p)}$. In order to apply Lemma \ref{lemma_search} we need to know the largest eigenvalue of  $A_{G(n,p)}$, which we denote as $\lambda^A_1$, its corresponding eigenstate $\ket{v_1}$ and the second largest eigenvalue of $A_{G(n,p)}$ denoted as $\lambda^A_2$. It was shown in Ref.~\cite{furedi1981eigenvalues} that the highest eigenvalue, $\lambda^A_1$ is a random variable whose probability distribution converges to a Gaussian distribution with mean $np$ and standard deviation $\sqrt{p(1-p)}$, as $n\rightarrow \infty$. The corresponding eigenstate, $\ket{v_1}$ tends \emph{almost surely} to $\ket{s}=1/\sqrt{n}\sum_{i=1}^n \ket{i}$. For a more detailed analysis of the convergence of $\ket{v_1}$ to $\ket{s}$, refer to Lemma 2 in Section III of Supplemental Material. It is also possible to obtain an upper bound on the second highest eigenvalue, $\lambda^A_2$ from the results of Ref.~\cite{furedi1981eigenvalues} which applies to random symmetric matrices. In fact in Ref.~\cite{vu2007}, a tighter bound on $\lambda^A_2$ is provided as $n\rightarrow \infty$, given by
\begin{equation}
\lambda^A_2=2\sqrt{np}+\mathcal{O}((np)^{1/4}\log(n))
\end{equation}
We see that as long as $p\geq \log^{4/3}(n)/n$, the ratio $\lambda^A_2/\lambda^A_1$ is bounded by a constant. However, as can be seen in Section III of Supplemental Material, in order to ensure that $\ket{v_1}$ converges to $\ket{s}$, almost surely, we choose the critical value of probability for search to be optimal as $p\geq \log^{3/2}(n)/n$. In fact, in the asymptotic limit, $\lambda^A_2/\lambda^A_1\rightarrow 0$, and the eigenstates corresponding to the two lowest eigenvalues of $H_{G(n,p)}$ are
\begin{equation}
\label{appev}
\ket{\lambda_{\pm}}\approx\frac{\ket{w}\pm\ket{s_{\bar{w}}}}{\sqrt{2}},
\end{equation}
where $\ket{s_{\bar{w}}}$ is the equal superposition of all the vertices other than the solution state $\ket{w}$. The probability of success is
\begin{equation}
\label{probsucc}
P_{w}(t)=|\braket{w|\exp(-i H_{G(n,p)} t)|s}|^2= \sin^2{\left(\frac{t}{\sqrt{n}}\right)}.
\end{equation} 
To confirm these theoretical predictions we plot, on the left side of Fig.~\ref{fig:probs_spectra} a)-c), the approximate probability $P_w(t)$ from Eq.~(\ref{probsucc}) (in red) and the exact solution calculated numerically (in blue) for $n=1000$ and $p=0.1, 0.01, 0.002$. On the right side, we plot the spectrum of the respective Hamiltonians. We observe, as expected, that the larger the gap between the two lowest eigenvalues and the bulk of the spectrum, the better is the approximation given by Eq.~(\ref{probsucc}) for the probability of success of search. As this gap disappears, close to the percolation threshold, the eigenstates corresponding to the two lowest eigenvalues do not follow Eq.~(\ref{appev}) and will mix randomly with the subspace orthogonal to $\ket{w}$ and $\ket{s_{\bar{w}}}$. At this point, since we are close to the percolation threshold, the graph is expected to have some isolated components and the algorithm breaks (see Fig.~\ref{fig:probs_spectra} c)). 

So far we have made the choice $\gamma=1/\lambda^A_1$, and assumed that we know the value of the random variable $\lambda^A_1$. In fact, its standard deviation is small enough so that it is sufficient to know its mean, which is equal to $np$, i.e we can choose $\gamma=1/(np)$, in order to prove that search is optimal almost surely. We prove this in Section IV of Supplemental Material, using tools of degenerate perturbation theory. These tools are also useful to design protocols for performing optimal state transfer and entanglement generation in Erd\"{o}s-Renyi graphs, as will be explained subsequently.
\begin{figure}[h!]
\centering
\includegraphics[scale=0.35]{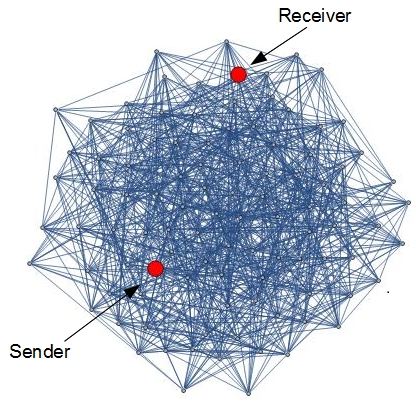}
\caption{\small{Quantum state transfer in Erd\"os-Renyi random graph $G(100,0.2)$: using our protocol, the fidelity achieved for this network is $80\%$.}}
\label{random-graph}
\end{figure}
\\
\emph{\textbf{State transfer with high fidelity --}} Quantum state transfer in spin chains \cite{bose} and spin networks \cite{datta} has been proposed as a way to establish short-range quantum channels. The problem of what structures lead to high fidelity state transfer has been of wide interest  \cite{datta,bose2,viv}. Here we show that it is possible to transfer, with low control and high fidelity, a quantum state between two arbitrary non-adjacent nodes of a random network (namely, an  Erd\"os-Renyi random graph). The Hamiltonian of a network of coupled spins, with an XX type interaction, conserves the number of excitations and so, in the  single excitation subspace, the Hamiltonian is that of a single particle quantum walk on the same network. The graph $G(n,p)$ can be perceived as a communication network where each node represents a party that transfers information to any of the other nodes. We assume that each party has access to a qubit and can control the local energy of the corresponding node. In order to transfer a state from node $i$ to $j$, with fidelity that tends to 1 in the asymptotic limit, the strategy is the following: all qubits are initially in state $\ket{0}$, which is an eigenstate of the network; the sender (corresponding to node $i$) and the receiver (corresponding to node $j$) can tune the respective site energies of $\ket{i}$ and $\ket{j}$ to $-1$, thereby making $\ket{i}$, $\ket{j}$ and $\ket{s}$ approximately degenerate. Finally, in order to transfer a qubit from $i$, the sender performs a local operation on her qubit to prepare $\ket{\psi}=\alpha\ket{0}+\beta\ket{1}$. As long as $p\geq\log^{3/2}(n)/n$, the approximate dynamics of a quantum walk starting at $\ket{i}$ is obtained by diagonalizing the Hamiltonian 
\begin{equation}
\label{state_transfer_Ham}
H'_{G(n,p)}=-\ket{i}\bra{i}-\ket{j}\bra{j}-\ket{s_{\bar{ij}}}\bra{s_{\bar{ij}}}-\gamma A'_{G(n,p)}
\end{equation}
projected onto the approximately degenerate subspace spanned by $\{\ket{i},\ket{s_{\bar{ij}}},\ket{j}\}$ which is given by 
\begin{equation}
H'_{G(n,p)}=\begin{bmatrix}
-1 && -1/\sqrt{n} && 0\\
-1/\sqrt{n}&&-1&&-1/\sqrt{n}\\
0 &&-1/\sqrt{n}&&-1
\end{bmatrix},
\label{ham_state}
\end{equation}  
with $\ket{s_{\bar{ij}}}=\sum_{k\neq i,j}\ket{k}/\sqrt{n-2}$ and $\ket{s_{\bar{ij}}}\approx \ket{s_{\bar{i}}}\approx \ket{s}$, where we assume that $i$ and $j$ are non-adjacent vertices. Thus, the dynamics is approximately the same as that of end-to-end state transfer in a chain with three spins, where perfect state transfer is possible \cite{viv} and the component of the wave function at the receiver is approximately $|\braket{j|U(t)|i}|^2=\sin^2(t/\sqrt{2n})$. Hence, after time $T=\pi\sqrt{n/2}$, the receiver gets $\ket{\psi}$ with fidelity 1, in the limit $n\rightarrow \infty$ (see Fig.~\ref{random-graph} for an example with finite $n$). The receiver can preserve this state for future use by tuning the energy of node $j$, locally, to a value that is off-resonant with the rest of the network \footnote{The error in this analysis is in going up to only first order in degenerate perturbation theory.}. We conclude that high fidelity quantum state transfer can be achieved in \textit{almost all} networks.
\\
\textit{\textbf{Creating Bell pairs in a random network --}} In quantum communication networks, entanglement is an useful resource that can be used for various tasks such as teleportation, superdense coding, cryptographic protocols, etc \cite{nielchuang}. Here, we present a protocol to entangle arbitrary nodes in a random network based on the search Hamiltonian. Imagine that Charlie at node $\ket{w}$ wants to entangle the qubits of Alice at node $\ket{a}$ and of Bob at node $\ket{b}$. We assume that none of the nodes $\ket{w}, \ket{a}$ and $\ket{b}$ are adjacent to each other. As before, $\gamma$ is chosen to be $1/(np)$. In this case, the protocol is as follows: i) Alice, Bob and Charlie tune their respective site energies to $-1$, ii) Charlie tunes his nearest neighbour couplings to $\sqrt{2}/d_C$, where $d_C$ is the degree of the node corresponding to Charlie, while the other couplings in the graph are $\gamma=1/np$. This ensures that the Hamiltonian, projected onto the approximately degenerate subspace spanned by $\ket{w}$, $\ket{s_{\overline{wab}}}=\sum_{k\neq a,b,w}\ket{k}/\sqrt{n-2}$ and $\ket{s_{ab}}=(\ket{a}+\ket{b})/\sqrt{2}$, is equal to 
\begin{equation}
H'_{G(n,p)}=\begin{bmatrix}
-1 && -\sqrt{2/n} && 0\\
-\sqrt{2/n}&&-1&&-\sqrt{2/n}\\
0 &&-\sqrt{2/n}&&-1
\end{bmatrix},
\label{ham_bell}
\end{equation} 
in the asymptotic limit \footnote{The state $\ket{s^-_{ab}}=(\ket{a}-\ket{b})/\sqrt{2}$ is also degenerate with these states since $\braket{s^-_{ab}|H|s^-_{ab}}\approx -1$. However, this state is decoupled from the dynamics because $\braket{s_{\overline{wab}}|H|s^-_{ab}}\approx -1$}. Thus, after time $T=\pi\sqrt{n}/2$, Alice and Bob share the state $\ket{s_{ab}}=(\ket{a}+\ket{b})/\sqrt{2}$, which is a Bell state. Subsequently, other Bell states may be obtained by local operations. Furthermore, Alice and Bob can preserve their Bell state by tuning the local energies of their qubits to a value that is off-resonant with the other eigenvalues of the network. 

\textbf{\textit{Discussion -- }} We have shown that searching for a marked node in a graph using continuous-time quantum walks works optimally for almost all graphs. This means that, in terms of the structures on which it performs optimally, this approach to quantum spatial search is much more general than what has been shown before. Our result was obtained by proving that the algorithm is \textit{almost surely} optimal for Erd\"os-Renyi random graphs $G(n,p)$, as long as $p\geq \log^{3/2}(n)/n$. 

As pointed out in Ref.~\citep{Childs_spatial_search}, the analog version of Grover's algorithm of Ref.~\citep{Farhi_analog_grover} can be seen as a quantum walk on the complete graph. Furthermore, Erd\"{o}s-Renyi random graph $G(n,p)$ can be obtained from the complete graph by randomly deleting edges with probability $1-p$. Thus, our result can also be interpreted as showing an inherent robustness of the analog version of Grover's algorithm to edge loss. This implies that there is a large family of random Hamiltonians that can be employed to achieve optimal quantum search. Hence, our work paves the way to understanding how this randomness would translate to the circuit model of quantum search and whether this implies an inherent robustness of the (standard) Grover's algorithm.

Finally, we have shown that one can adapt the spatial search algorithm to design  protocols for quantum state transfer and for entanglement generation between arbitrary nodes of a random network of interacting qubits. Our results show  that quantum information tasks typically designed for structured systems retain performance in very disordered structures. These results could lead to further investigation on what kind of random structures appear naturally in physical systems (for example those appearing in Refs.~\cite{masoud, buchleitner}) and whether they would offer a sufficient spectral gap to perform efficient and robust quantum information tasks. It would also be interesting to explore whether non-trivial quantum information tasks can be performed on other models of random networks such as scale-free networks \cite{newman}.

\textbf{\textit{Acknowledgements -- }} LN, SC and YO thank the support from Funda\c{c}\~{a}o para a Ci\^{e}ncia e a Tecnologia (Portugal), namely through programmes PTDC/POPH/POCH and projects UID/EEA/50008/2013, IT/QuSim, ProQuNet, partially funded by EU FEDER, and from the EU FP7 projects  LANDAUER (GA 318287) and PAPETS (GA 323901). Furthermore, LN and SC acknowledge the support from the DP-PMI and FCT (Portugal) through SFRH/BD/52241/2013 and SFRH/BD/52246/2013, respectively. AA thanks the support from ERC Advanced Grant MQC (320731), EU FP7 projects QALGO (600700) and RAQUEL (323970), and the Latvian State Research Programme NexIT Project No.~1.

\section*{Supplemental Material}
\section{Sufficient condition for optimal quantum search: Proof of Lemma 1}
\proof Let us express $\ket{w}$ in the basis $\ket{v_1}, \ket{v_2}, \ldots, \ket{v_k}$:  
\begin{equation}
\label{sol}
\ket{w} = a_1 \ket{v_1} + a_2 \ket{v_2} + \ldots + a_k \ket{v_k}.
\end{equation} 
We rescale $H_1$ by $(1+r)H_1-rI$, where $I$ is the identity matrix. With this replacement, $\ket{v_1}$ remains an eigenvector of $H_1$ with an eigenvalue 1. The other eigenvalues change to $\lambda'_i=(1+r)\lambda_i - r$. Now,  the expression 
\begin{equation}
\sum_{i=2}^{k}\frac{a_i^2}{1-\lambda_i},
\end{equation}
after rescaling $H_1$ as mentioned before becomes
\begin{align}
\sum_{i=2}^{k}\frac{a_i^2}{1-\lambda'_i}&=\sum_{i=2}^{k}\frac{a_i^2}{(1+r)(1-\lambda_i)}.
\end{align}

As $|\lambda_i|\leq c$, we have
\begin{align}
&\sum_{i=2}^{k}\frac{a_i^2}{(1+r)(1+c)}\leq\sum_{i=2}^{k}\frac{a_i^2}{(1+r)(1-\lambda_i)}, \text{and,}\\ &\sum_{i=2}^{k}\frac{a_i^2}{(1+r)(1-\lambda_i)}\leq \sum_{i=2}^{k}\frac{a_i^2}{(1+r)(1-c)}
\end{align}

So we choose an appropriate $r\in[-\frac{c}{1+c},\frac{c}{1-c}]$, such that 
\begin{equation}
\label{eq:rescale}
\sum_{i=2}^k \frac{a_i^2}{1-\lambda_i} = \sum_{i=2}^k a_i^2.
\end{equation}
If $|\lambda_i|\leq c$, then $\lambda'_i \geq 0$ for $r=-\frac{c}{1+c}$ and
$\lambda'_i \leq 0$ for $r=\frac{c}{1-c}$.
In the first case, the left hand side of (\ref{eq:rescale}) 
is at least the right hand side. In the second case, the left hand side is at most the right hand side.
After the replacement of $H_1$ by $(1+r)H_1-rI$, the new eigenvalues $\lambda'_2, \ldots, \lambda'_k$
are in the interval $[-\frac{2c}{1-c}, \frac{2c}{1+c}]$.

After we replace $H_1+H_2$ by $(1+r)H_1+H_2-rI$, we can omit the $-rI$ term (since it only
affects the phase of the state). To simplify the notation, we now refer to the Hamiltonian $(1+r)H_1$ as
$H_1$ and to new eigenvalues $\lambda'_i$ as $\lambda_i$. 

Let 
\begin{equation} 
\ket{v} = b_1 \ket{v_1} + b_2 \ket{v_2} + \ldots + b_k \ket{v_k}.
\end{equation}
We write out the conditions for $\ket{v}$ to be an eigenvector of $H_1+H_2$
with an eigenvalue $\lambda$.
We have
\begin{equation}
H\ket{v} = H_1\ket{v}+H_2\ket{v} = \sum_i (b_i\lambda_i+a_i \gamma) \ket{v_i}
\end{equation}
where $\gamma = \lbra w \ket{v}$. Since we also have
\begin{equation}
 H\ket{v} =\lambda \ket{v} = \sum_i \lambda b_i \ket{v_i},
\end{equation}
we get that $\lambda b_i = \lambda_i b_i + \gamma a_i$ which is equivalent to
$b_i = \frac{\gamma}{\lambda - \lambda_i} a_i$. Substituting this into
$\gamma = \lbra w \ket{v} = \sum_i a_i b_i$ gives that 
\begin{equation}
\sum_i \frac{a_i^2}{\lambda - \lambda_i} = 1.
\end{equation}
This is the condition for the eigenvalues $\lambda$. In each of intervals $[\lambda_i, 
\lambda_{i-1}]$ for $i=2, \ldots, k$, the left hand side
is strictly decreasing from $+\infty$ to $-\infty$
and in the interval $[\lambda_1, +\infty)$, the left hand side is 
strictly decreasing from $+\infty$ to 0. Therefore, each of these intervals 
contains one eigenvalue.

We are interested in the two eigenvalues $\lambda$ that are in $[\lambda_2, \lambda_1]$ 
and $[\lambda_1, +\infty)$. (We denote these eigenvalues by $\lambda_-$ and $\lambda_+$ 
and the corresponding eigenvectors by $\ket{v_-}$ and $\ket{v_+}$.) We express these 
eigenvalues as $\lambda = 1+\delta$ (where $\delta$ is positive for $\lambda_+$ and negative for $\lambda_-$). 
By Taylor expansion, if $\delta$ is small, we have
\begin{equation}
\sum_{i=2}^k \frac{a_i^2}{(1+\delta)-\lambda_i} =  \sum_{i=2}^k \frac{a_i^2}{1-
\lambda_i} - \sum_{i=2}^k \frac{a_i^2}{(1-\lambda_i)^2} \delta + O(\delta^2) .
\end{equation}
Thus, the condition for eigenvalues becomes 
\begin{equation}
\frac{a_1^2}{\delta} + \sum_{i=2}^k \frac{a_i^2}{1-\lambda_i} - \sum_{i=2}^k \frac
{a_i^2}{(1-\lambda_i)^2} \delta + O(\delta^2) = 1 .
\end{equation}
Since the second term on the left hand side is 1, this is equivalent to
\begin{equation}
\frac{a_1^2}{\delta} = \sum_{i=2}^k \frac{a_i^2}{(1-\lambda_i)^2} \delta + O
(\delta^2).
\end{equation}
which is satisfied for 
\begin{equation}
\label{eq:delta} 
\delta \approx \pm \frac{a_1}{\sqrt{\sum_{i=2}^k \frac{a_i^2}{(1-\lambda_i)^2}}} .
\end{equation}
Since $a_1=\epsilon$ and the denominator is of the order $\Theta(1)$, the right hand side is $\Theta
(\epsilon)$.

We now consider the overlap $\lbra s\ket{v_+}$. We assume that $\ket{v_+}$ is 
normalized so that $\|v_+\|=1$. This is equivalent to $\sum_i b_i^2 = 1$ which, in turn, is equivalent to 
\begin{equation}
 \sum_i \left( \frac{\gamma}{\lambda - \lambda_i} a_i \right)^2 =1 .
\end{equation}
We can rewrite this as
\begin{equation}
 \frac{1}{\gamma} = \sqrt{\sum_i \frac{a_i^2}{(\lambda-\lambda_i)^2}} .
\end{equation}
We now estimate the expression under the square root. We have
\begin{equation}
\label{eq:gamma1} 
\frac{1}{\gamma} \approx \sqrt{\frac{a_1^2}{\delta^2} + \sum_{i=2}^k \frac{a_i^2}{(1-\lambda_i)^2}}  
\approx \sqrt{2} \frac{a_1}{\delta} 
\end{equation}
with the first step following by approximating $\lambda-\lambda_i = (1+\delta)-\lambda_i \approx 1-\lambda_i$
for $i>1$ and the second step follows from (\ref{eq:delta}).

This means that $\gamma \approx \frac{\delta}{\sqrt{2} a_1}$.
Therefore,
\begin{equation}
 \lbra s \ket{v_+} = \frac{\gamma a_1}{\lambda_+-1} = \frac{\gamma a_1}{\delta} 
\approx \frac{1}{\sqrt{2}}
\end{equation}
and
\begin{equation}
\lbra s\ket{v_-} = \frac{\gamma a_1}{\lambda_--1} = -\frac{\gamma a_1}{\delta} 
\approx -\frac{1}{\sqrt{2}} .
\end{equation}
Thus, $\ket{s}$ can be approximated by $\frac{1}{\sqrt{2}} (\ket{v_+}- \ket{v_-})$.
Evolving the Hamiltonian $H_1+H_2$ for time $\frac{\pi}{2\delta} = \Theta(\frac{1}{\epsilon})$
transforms $\ket{s}$ to 
\begin{equation}
\ket{f} \approx \frac{1}{\sqrt{2}} (\ket{v_+}+ \ket{v_-}).
\end{equation}
We have 
\begin{equation}
\label{eq:w} \lbra w \ket{f} \approx \frac{1}{\sqrt{2}} \lbra w \ket{v_+} +  
\frac{1}{\sqrt{2}} \lbra w \ket{v_-} .
\end{equation}
We now consider $\lbra w \ket{v_+} = \gamma$.
By combining the first part of (\ref{eq:gamma1}) with (\ref{eq:delta}), we obtain that 
\begin{equation}
\frac{1}{\gamma} \approx \sqrt{2 \sum_{i=2}^k \frac{a_i^2}{(1-\lambda_i)^2}} .
\end{equation}
Because of (\ref{eq:rescale}) and $\lambda_i \leq \frac{2c}{1+c}$, this is at most 
\begin{equation}
 \sqrt{2 \frac{\sum_{i=2}^k a_i^2}{1-\frac{2c}{1+c}}} \leq \sqrt{2 \frac{1}{1-\frac{2c}{1+c}}} = \sqrt{\frac{2(1+c)}{1-c}}.
\end{equation}
Therefore, $\gamma = \lbra w\ket{v_+} \geq \sqrt{\frac{1-c}{2(1+c)}}$. 
Similarly, $\lbra w\ket{v_-}\geq \sqrt{\frac{1-c}{2(1+c)}}$. 
Together with (\ref{eq:w}), this means that, up to the approximations that we made,
\begin{equation}
\lbra w \ket{f} \geq \sqrt{\frac{1-c}{1+c}}. 
\end{equation}
\hfill\qed
\section{Quantum search on random regular graphs}
A family of random graphs whose adjacency matrix has an $\mathcal{O}(1)$ gap between the largest and second largest eigenvalues are the $d$-random regular graphs, a random graph sampled uniformly from the set of all regular graphs of degree $d$. For these graphs, the largest eigenvalue is $d$, with the corresponding eigenvector, $\ket{s}$. Also in Ref.~\cite{broder1987second} it has been proven that the second largest eigenvalue is $\mathcal{O}(d^{3/4})$ for $d\geq 3$, with high probability. This way, we choose $\gamma=1/d$ and since $\lambda_2^A/\lambda_1^A=\mathcal{O}(d^{-1/4})<1$, it follows from Lemma \ref{lemma_search} that quantum search is optimal. It is interesting to note that for lattices, the lowest dimension for which search is possible in $\mathcal{O}(\sqrt{n})$ time is dimension five \citep{Childs_spatial_search}, which is a specific instance of a regular graph of degree ten. However, for random regular graphs search is optimal for degree three and larger. 
\section{Convergence of the eigenstate corresponding to the maximum eigenvalue of an Erd\"{o}s-Renyi random graph}
\begin{Lemma}
\label{convergence_lemma}
Let $A$ be the adjacency matrix of the Erd\"os-Renyi random graph $G(n,p)$ with vertices $1,2,...,n$. Let $\gamma A$ represent the adjacency matrix of $G(n,p)$ with each entry rescaled by $\gamma=1/np$. Also let $\ket{s}=(1/\sqrt{n})\sum_{i=1}^{n}\ket{i}$ be the equal superposition of all nodes such that $\ket{s}=\alpha\ket{v_1}+\beta\ket{v_1}^{\perp}$, where $\ket{v_1}$ is the eigenvector corresponding to the highest eigenvalue, $\lambda_1$ of $\gamma A$. Then, $\alpha\geq 1-o(1)$ almost surely for $p\geq \frac{\log^{3/2} n}{n}$.
\end{Lemma}
\proof
First we observe that from Ref.~\cite{furedi1981eigenvalues} that the largest eigenvalue follows a Gaussian distribution with mean $1$ and standard deviation $\frac{1}{n}\sqrt{\frac{1-p}{p}}$, i.e,  $\lambda_1\sim\mathcal{N}(1,\frac{1}{n}\sqrt{\frac{1-p}{p}})$. So if $\delta=1/\sqrt{n}$, for $p\geq \log^{3/2} n/n$, one can show that 
\begin{equation}
Pr[\lambda_1\geq 1-\delta]=1-\frac{1}{2}\text{erfc}\Big{[}\frac{\log^{3/4} n}{\sqrt{2}}\Big{]},
\end{equation}
where $\text{erfc}[x]=(2/\sqrt{\pi})\int_{x}^{\infty}e^{-x^2/2}dx$. As $\text{erfc}[x]\rightarrow 0$ as $n\rightarrow \infty$, we have 
\begin{equation}
\lambda_1\geq 1-\delta,
\end{equation}
almost surely. To prove this explicitly, we use the bound,$ \text{ erfc}[x]\leq \frac{2}{\sqrt{\pi}}\frac{e^{-x^2}}{(x+\sqrt{x^2+4/\pi})}$, for $x>0$. In our case $x=(\log^{3/4} n)/\sqrt{2}$ and so, by using the inequality $\log^a n\geq a\log n$, for $a>1$, we can show that  
\begin{equation}
\text{erfc}[x]\leq \mathcal{O}\Big{(}\frac{1}{n^{3/4}\log^{3/4} n}\Big{)}.
\end{equation}
Thus,
\begin{equation}
\label{evaluebound}
Pr[\lambda_1\geq 1-\delta]\geq 1-\mathcal{O}\Big{(}\frac{1}{n^{3/4}\log^{3/4} n}\Big{)}.
\end{equation}
Similarly, one can also obtain an upper bound $\lambda_1\leq 1+\delta$, almost surely.  

Also, from Ref.~\citep{furedi1981eigenvalues, vu2007}, we have,
\begin{equation}
\label{boundnorm1}
||\gamma(A-E(A))||\leq (2+(np)^{-1/4}\log n)\frac{1}{\sqrt{np}},
\end{equation}
where $E(X)$ denotes the expectation of random variable $X$ and $||~||$ denotes the standard Euclidean norm.

Let $\lambda_i, j\geq 2$ be the rest of the spectrum of $\gamma A$ and let $\ket{v_i}$ be the corresponding eigenvectors.
From Eq.~\ref{boundnorm1}, it follows that
\begin{align}
\label{boundnorm2}
&||\lambda_1\ket{v_1}\bra{v_1}+\sum_{i\geq 2}\lambda_i\ket{v_i}\bra{v_i}-\ket{s}\bra{s}||\\
&\leq (2+(np)^{-1/4}\log n)\frac{1}{\sqrt{np}}
\end{align}
Now,
\begin{align}
&\Big{(}\lambda_1\ket{v_1}\bra{v_1}+\sum_{i\geq 2}\lambda_i\ket{v_i}\bra{v_i}-\ket{s}\bra{s}\Big{)}\ket{v_1}\\
&= \lambda_1\ket{v_1}-\alpha\ket{s}\\
&=(\lambda_1-\alpha^2)\ket{v_1}-\alpha\beta\ket{v_1}^{\perp}.
\end{align}
So,
\begin{align}
&||\Big{(}\lambda_1\ket{v_1}\bra{v_1}+\sum_{i\geq 2}\lambda_i\ket{v_i}\bra{v_i}-\ket{s}\bra{s}\Big{)}\ket{v_1}||^2\\
&=(\lambda_1-\alpha^2)^2+\alpha^2\beta^2.
\end{align}
From Eq.~\ref{boundnorm2} and the Cauchy-Schwarz inequality, we obtain
\begin{align}
&(\lambda_1-\alpha^2)^2\leq \frac{(2+(np)^{-1/4}\log n)^2}{np}\\
\implies&\alpha^2\geq \lambda_1-\frac{(2+(np)^{-1/4}\log n)}{\sqrt{np}}\\
\implies &\alpha\geq\alpha^2\geq 1-o(1),
\end{align}
for $p\geq \frac{\log^{3/2}n}{n}$, 
where the final expression for $\alpha$ follows from the fact that $\alpha\in [0,1]$ and that $\lambda_1\geq 1-\delta$.
\\
\begin{flushright}\qed\end{flushright}


\section{Proof of optimality of search on Erd\"{o}s-Renyi random graphs using degenerate perturbation theory}
Here, we present an intuitive way to prove the optimality of search for Erd\"{o}s-Renyi graphs, $G(n,p)$, by using degenerate perturbation theory. This proof is instrumental in constructing protocols for optimal state transfer and entanglement generation in these random networks. 

The spectral density of a graph G is defined as 
\begin{equation}
\rho(\lambda)=\frac{1}{n}\sum_{i=1}^n \delta(\lambda_i-\lambda)
\end{equation}
where $\lambda_i$ are eigenvalues of the adjacency matrix $A_G$ \cite{barabasi_review}. In the limit of $n\rightarrow\infty$ this approaches a continuous function. For a random graph $G(n,p)$, as long as $np\rightarrow\infty$, the spectral density is given by 
\begin{equation}\label{semicircle_law}
\rho(\lambda) = \begin{cases} \dfrac{\sqrt{4 n p(1-p)-\lambda^2}}{2\pi n p(1-p)} &\mbox{if } |\lambda|<2\sqrt{np(1-p)} \\
0 & \mbox{otherwise }\end{cases}, 	  
\end{equation}
known as the Wigner's semicircle law. The highest eigenvalue, $\lambda_1$, of $A_{G(n,p)}$ is isolated from the bulk of the spectrum and follows a Gaussian distribution with mean $np$ and standard deviation $\sqrt{p(1-p)}$, as $n\rightarrow \infty$. From Section I, the corresponding eigenstate, $\ket{v_1}$ tends \emph{almost surely} to $\ket{s}=1/\sqrt{n}\sum_{i=1}^n \ket{i}$.  From Ref.~\cite{vu2007}, we obtain that the second highest eigenvalue, as $n\rightarrow \infty$, is given by 
\begin{equation}
\lambda_2=2\sqrt{np}+\mathcal{O}((np)^{1/4}\log(n)).
\end{equation}

There is thus a significant gap in the spectrum between the first and second largest eigenvalues, with high probability. In order to make use of this separation between the largest eigenvalue and the rest of the spectrum, it will be helpful to write
\begin{equation}
A_{G(n,p)}= E \ket{s'}\bra{s'} + A'_{G(n,p)}
\end{equation}
where $E\rightarrow np$, and $\ket{s'}\rightarrow \ket{s}$ as $n\rightarrow \infty$. The search Hamiltonian then becomes 
\begin{equation}
H_{G(n,p)}=-\ket{w}\bra{w}-\gamma_p E \ket{s'}\bra{s'} -\gamma_p  A'_{G(n,p)}.
\end{equation}
We use $\ket{s}$ as the initial state of the quantum algorithm and choose $\gamma_p=1/(np)$ so that, for large $n$, $\ket{s}$ and $\ket{w}$ are approximately degenerate. The spectrum of $\gamma_p A'_{G(n,p)}$ follows the semi-circle law, where the radius of the semicircle is given by 
\begin{equation}\label{radius_Wigner}
R=\gamma_p 2\sqrt{n p (1-p)}=2\sqrt{\frac{(1-p)}{np}}. 
\end{equation}
As long as $np\rightarrow\infty$, the radius $R\rightarrow 0$. This implies, for the whole range in which the semi-circle law is valid, that the radius $R$ shrinks as $np$ grows. Also from Section III, we know that as long as $p\geq \log^{3/2}(n)/n$,
\begin{equation}
1-\frac{1}{\sqrt{n}}\leq\gamma_p\lambda_1\leq 1+\frac{1}{\sqrt{n}},
\end{equation}
almost surely. 

One can show that the algorithm retains its optimality as long as this is the case. If 
\begin{equation}
-1-\delta\leq\gamma_p\lambda_1\leq -1+\delta, 
 \end{equation}
using degenerate perturbation theory, we obtain the ground and first excited states of $H_{G(n,p)}$ from its diagonalization in the 2-dimensional subspace spanned by $\{\ket{w},\ket{s_{\bar{w}}}\}$, where $\ket{s_{\bar{w}}}=\sum_{i\neq w}^n \ket{i}/\sqrt{n-1}$. These eigenstates are
\begin{align}
\label{eigenstates_prediction}
\ket{\lambda_{+}}&\approx \frac{1}{\kappa}\Big{(}\frac{1}{\sqrt{n}}\ket{w}-\mu\ket{s_{\bar{w}}}\Big{)},\\
\ket{\lambda_{-}}&\approx \frac{1}{\kappa}\Big{(}\mu\ket{w}+\frac{1}{\sqrt{n}}\ket{s_{\bar{w}}}\Big{)}
\end{align}
where $\ket{s_{\bar{w}}}=\sum_{i\neq w}^n \ket{i}/\sqrt{n-1}$, $\mu=\delta/2+\sqrt{\delta^2/4+1/n}$ and $\kappa=\sqrt{\mu^2+1/n}$.

Thus, the probability of observing the marked vertex $\ket{w}$ is given by 
\begin{equation}\label{prediction}
P_{w}(t)=|\braket{w|\exp(-i H_{G(n,p)} t)|s}|^2= \frac{1}{1+N\delta^2/4}\sin^2{\left(\frac{\Omega t}{2}\right)},
\end{equation}
where $\Omega=\sqrt{\delta^2/4+1/n}$. Thus, as long as $\delta\in[-1/\sqrt{n},1/\sqrt{n}]$, the running time of the algorithm is $T=\mathcal{O}(\sqrt{n})$.

Thus, the gap between the lowest eigenvalue of $-\gamma_p A'_{G(n,p)}$ and the second lowest eigenvalue of $H_{G(n,p)}$, is $\gamma_p(\lambda_2-\lambda_1)=\mathcal{O}(1)$ as long as $p\geq\log^{3/2}(n)/n$. The error obtained from this approximation is $\mathcal{O}(\gamma_p^2 \lambda_2^2)$, since $\gamma_p \lambda_2$ is the largest eigenvalue of $\gamma_p A'_{G(n,p)}$ which we consider as a perturbation. Note that the error of our approximation decreases as $np$ increases. Thus, for a fixed $n$, the higher the value of $p$, the lower is the error. Again, for a fixed $p$, the error diminishes with increase in $n$.
\bibliographystyle{unsrt}
\bibliography{Bibliography}
\end{document}